\documentclass[11pt]{article}
\usepackage{amssymb}
\usepackage{amsmath}
\usepackage{amsthm}
\usepackage[dvips]{graphics,color}
\usepackage{epsfig}

\usepackage{a4wide}

\def\E{\mathbb{E}}

\newtheorem{thm}{Theorem}

\newtheorem{alg}{Algorithm}

\begin{document}

\title{Covariance Estimation for Multivariate Conditionally Gaussian Dynamic Linear Models}
\author{K. Triantafyllopoulos\footnote{Department of Probability and Statistics, University of Sheffield,
UK,
email: \tt{k.triantafyllopoulos@sheffield.ac.uk}}}

\date{1 March 2006}

\maketitle

\begin{abstract}

In multivariate time series, the estimation of the covariance matrix
of the observation innovations plays an important role in
forecasting as it enables the computation of the standardized
forecast error vectors as well as it enables the computation of
confidence bounds of the forecasts. We develop an on-line,
non-iterative Bayesian algorithm for estimation and forecasting. It
is empirically found that, for a range of simulated time series, the
proposed covariance estimator has good performance converging to the
true values of the unknown observation covariance matrix. Over a
simulated time series, the new method approximates the correct
estimates, produced by a non-sequential Monte Carlo simulation
procedure, which is used here as the gold standard. The special, but
important, vector autoregressive (VAR) and time-varying VAR models
are illustrated by considering London metal exchange data consisting
of spot prices of aluminium, copper, lead and zinc.

\textit{Some key words:} Multivariate time series, dynamic linear
model, Kalman filter, vector autoregressive model, London metal
exchange.

\end{abstract}

\section*{Introduction}

Multivariate time series receive considerable attention because a
great deal of time series data arrive in vector form. Whittle (1984)
and L\"utkepohl (1993) discuss VARMA models for vector responses,
whilst Harvey (1989, Chapter 8), West and Harrison (1997, Chapter
16) and Durbin and Koopman (2001, Chapter 3) extend this work to
state space models for observation vectors. In econometrics most
studies of state space models focus on trend estimation, signal
extraction and volatility. A review of recent developments of state
space models in econometrics can be found in Pollock (2003). Barassi
{\it et al.} (2005) and Gravelle and Morley (2005) give applications
of the Kalman filter to interest rates data and Harvey {\it et al.}
(1994) use Kalman filter techniques to estimate the volatility of
foreign exchange rates using multivariate stochastic volatility
(MSV) models. With the exception of multivariate GARCH and MSV
models, which focus on the prediction of the volatility, it is
usually desirable to use a structural state space model to forecast
time series vectors (e.g. foreign exchange rates, monthly sales,
interest rates, etc) and to estimate the observation innovation
covariance matrix of the underlying time series. For such
applications and for short term forecasting the above covariance
matrix can be assumed time-invariant, but unknown, and its
estimation is the main aim of this paper.

The estimation of the observation covariance matrix plays an
important role in forecasting. Firstly we note that, under the
general multivariate dynamic linear model (see equation
(\ref{model1}) below), the multi-step forecast mean of the
response time series vector is a non-linear function of the
observation covariance matrix (West and Harrison, 1997, Chapter
16). Secondly, the computation of the standardized forecast error
vectors requires a precise estimation of the observation
covariance matrix and thus a miss-specification of the observation
covariance matrix can lead to false results regarding the
evaluation and judgement of the model. Thirdly, the multi-step
forecast covariance matrix is a linear function of the observation
covariance matrix and the former is of particular interest; the
forecast covariance matrix can explain the variability of the
forecasts and hence it can enable the computation of confidence
bounds for the forecasts. Finally, the precise estimation of the
observation covariance matrix gives an accurate estimation of the
cross-correlation structure of the several component time series,
which is particularly useful, especially for financial time
series. For all the above reasons the study of the estimation of
the observation covariance matrix is worthwhile and its
contribution to forecasting for multivariate time series is
paramount.

The problem of the estimation of the observation innovation
variance for univariate state space models has been well reported
(West and Harrison, 1997, \S4.5; Durbin and Koopman, 2001,
\S2.10), however, for vector time series this problem becomes
considerably more complex and the available methodology consists
of special cases, approximations and iterative procedures.

Let $y_t$ be a $p$-dimensional observation vector following the
Gaussian dynamic linear model (DLM):
\begin{equation}\label{model1}
y_t=F'\theta_t+\epsilon_t\quad \textrm{and} \quad \theta_t = G
\theta_{t-1}+\omega_t,
\end{equation}
where $\theta_t$ is a $d$-dimensional Markovian state vector, $F$ is
a known $d\times p$ design matrix and $G$ a known $d\times d$
transition matrix. The notation $F'$ is used for the transpose
matrix of $F$. The distributions usually adopted for
$\{\epsilon_t\}$, $\{\omega_t\}$ and $\theta_0$ are the multivariate
Gaussian, i.e. $\epsilon_t\sim \mathcal{N}_p(0,\Sigma)$,
$\omega_t\sim \mathcal{N}_d(0,\Omega)$ and $\theta_0\sim
\mathcal{N}_d(m_0,P_0)$, for some known priors $m_0$ and $P_0$. The
innovation vectors $\{\epsilon_t\}$ and $\{\omega_t\}$ are assumed
individually and mutually uncorrelated and they are also assumed
uncorrelated with the initial state vector $\theta_0$, i.e. for all
$t\neq s$: $\E(\epsilon_t\epsilon_s')=0$, $\E(\omega_t\omega_s')=0$,
and for all $t,s>0$: $\E(\epsilon_t\omega_s')=0$,
$\E(\epsilon_t\theta_0')=0$ and $\E(\omega_t\theta_0')=0$, where
$\E(\cdot)$ denotes expectation. The covariance matrices $\Sigma$
and $\Omega$ are typically unknown and their estimation or
specification is a well known problem. The interest is centered on
the estimation of $\Sigma$, while $\Omega$ can be specified \emph{a
priori} (West and Harrison, 1997, Chapter 6; Durbin and Koopman,
2001, \S 3.2.2).

Several methods have been proposed, for the estimation of $\Sigma$.
Harvey (1986) and Quintana and West (1987) independently introduce
matrix-variate DLMs, which are matrix-variate linear state space
models allowing for covariance estimation. Harvey (1986) proposes a
likelihood estimator, while Quintana and West (1987) propose a
Bayesian estimation modelling $\Sigma$ with an inverted Wishart
distribution. Harvey (1986)'s model is reported and further
developed in Harvey (1989), Fern\'{a}ndez and Harvey (1990), Harvey
and Koopman (1997) and Moauro and Savio (2005), while Quintana and
West (1987)'s model is reported and further developed in Quintana
and West (1988), Queen and Smith (1992), West and Harrison (1997),
Salvador {\it et al.} (2003), Salvador and Gargalo (2004) and
Salvador {\it et al.} (2004). However, both suggestions (Harvey
(1989)'s and Quintana and West (1987)'s) are criticized in Barbosa
and Harrison (1992) where it is shown that the above models are
restrictive in the sense that one can decompose the response vector
$y_t$ into several scalar time series and model each of these time
series individually, using univariate DLMs. Barbosa and Harrison
(1992) propose an approximate algorithm for the general DLM
(\ref{model1}), but their main assumption seems rather unjustified,
since it suggests that for any $p\times p$ matrix $C$ it is
$\Sigma^{1/2}C\Sigma^{-1/2}=\widehat{\Sigma}^{1/2}C\widehat{\Sigma}^{-1/2}$,
where $\widehat{\Sigma}$ is a point estimate of $\Sigma$ and the
notation $\Sigma^{1/2}$ stands for the symmetric square root of
$\Sigma$ (Gupta and Nagar, 1999, p. 7). This assumption holds
clearly when $\widehat{\Sigma}^{1/2}$, $C$ commute and when
$\Sigma^{1/2}=\Sigma^*$, $C$ commute, where $(\Sigma^*)^2$ is any
particular realization of $\Sigma$. However, in general the above
assumption is difficult to check since $\Sigma$ is the unknown
covariance matrix subject to estimation. In addition, that
assumption seems to be probabilistically quite inappropriate, since
it translates that the non-stochastic quantity
$\widehat{\Sigma}^{1/2}C\widehat{\Sigma}^{-1/2}$ equals the
stochastic quantity $\Sigma^{1/2}C\Sigma^{-1/2}$ with probability 1.
A possible analysis can be obtained in special cases where $\Sigma$
is diagonal or when the off-diagonal elements of $\Sigma$ are all
common. Triantafyllopoulos and Pikoulas (2002) and
Triantafyllopoulos (2006) adopt the model of Harvey (1986) and they
provide an improved on-line estimator for $\Sigma$ based on a
standard maximum likelihood technique. The problem is again that the
models discussed lack the general formulation of the state space
model (\ref{model1}); e.g. one can easily show that all above models
are special cases of model (\ref{model1}). Iterative procedures via
maximum likelihood and Markov chain Monte Carlo (MCMC) techniques
are available, but they tend to be slow, especially as the dimension
of the observation vector $p$ increases. Kitagawa and Gersch (1996),
Shumway and Stoffer (2000, Chapter 4), Durbin and Koopman (2001,
Chapter 7) and Doucet {\it et al.} (2001) discuss univariate
modelling with iterative methods, but their efficiency in
multivariate time series is not yet explored. Barbosa and Harrison
(1992) and West and Harrison (1997, \S 16.2.3) discuss the problem
of inefficiency of iterative methods and they point out that the
number of parameters to be estimated in $\Sigma$ is $p(p+1)/2$,
which rapidly increases with the dimension $p$ of the response
vector, e.g. for $p=10$ there are 55 distinct parameters in $\Sigma$
to be estimated. In addition to this Dickey {\it et al.} (1986)
discuss relevant issues on specifying and assessing the prior
distribution of $\Sigma$ pointing out difficulties in the
implementation of iterative procedures.

In this paper we propose a new non-iterative Bayesian procedure for
estimating $\Sigma$ and for forecasting $y_t$. This procedure offers
a novel estimator of $\Sigma$ for the general DLM (\ref{model1}).
The proposed estimator is empirically found to converge to the true
value of $\Sigma$ and this estimator approximates well the
respective estimators in the special cases of the conjugate
univariate and matrix-variate DLMs. A comparison with a
non-sequential Monte Carlo simulation shows that the new method
produces estimates close to the MCMC. The focus and the benefit
employing the new method is on on-line estimation and therefore no
attempt has been made to compare the proposed algorithms with
sequential iterative procedures. The reason for this is justified by
the above discussion and the interested reader should refer to
Dickey {\it et al.} (1986) and West and Harrison (1997, \S 16.2.3).
The proposed forecasting procedure for model (\ref{model1}) is
applied to the important model subclasses of vector autoregressive
(VAR) and VAR with time-dependent parameters. These models are
illustrated by considering London metal exchange data, consisting of
spot prices of aluminium, copper, lead and zinc (Watkins and
McAleer, 2004).

We begin by developing the main idea of the paper and giving the
proposed algorithm. The performance of this algorithm is illustrated
in the following section by considering simulated time series data;
a comparison with a Monte Carlo simulation is performed. The
proceeding section gives an application to vector autoregressive
modelling, which is used to analyze London metal exchange data, in
the following section. The appendix details a proof of a theorem in
the paper and it describes the MCMC simulation procedure.

\section*{Main Results}

Denote with $y^t=(y_1,y_2,\ldots,y_t)$ the information set
comprising data up to time $t$, for some positive integer $t>0$. Let
$m_t$ and $P_t$ be the posterior mean and covariance matrix of
$\theta_t|y^t$ and $S_t$ be the posterior expectation of $\Sigma$,
i.e. $\E(\Sigma |y^t)=S_t$. Let $y_t(1)=\E(y_{t+1}|y^t)=F'Gm_t$ be
the one-step forecast mean at time $t$ and
$Q_{t+1}=\textrm{Var}(y_{t+1}|y^t) =F'R_{t+1}F+S_t$ be the one-step
forecast covariance matrix at $t$, where $R_{t+1}=GP_tG'+\Omega$.
Upon observing $y_{t+1}$, we define the one-step forecast error
vector as $e_{t+1}=y_{t+1}-y_t(1)$. The next result (proved in the
appendix) gives an approximate property of $S_t$.
\begin{thm}\label{th1}
Consider the dynamic linear model (\ref{model1}). Let $\Sigma$ be
the covariance matrix of the observation innovation $\epsilon_t$ and
assume that $\lim_{t\rightarrow\infty}S_t=\Sigma$, where
$\E(\Sigma|y^t)=S_t$ is the true posterior mean of $\Sigma$ given
$y^t$. Let $n_0$ be a positive scalar and $S_0=\E(\Sigma)$ be the
prior expectation of $\Sigma$. If $\Sigma$ is bounded, then for
large $t$ the following holds approximately
\begin{equation}\label{eq3}
S_t=\frac{1}{n_0+t}\left(n_0S_0+\sum_{i=1}^tS_{i-1}^{1/2}Q_i^{-1/2}
e_ie_i'Q_i^{-1/2}S_{i-1}^{1/2}\right),
\end{equation}
where $e_i$, $Q_i$ are defined above and $S_{i-1}^{1/2}$,
$Q_i^{-1/2}$ denote respectively the symmetric square roots of the
matrices $S_{i-1}$, $Q_i^{-1}$ based on the spectral decomposition
factorization of symmetric positive definite matrices
$(i=1,2,\ldots,t)$.
\end{thm}
Conditionally now on $\Sigma=S$, for a particular value $S$, we can
apply the Kalman filter to the DLM (\ref{model1}) and obtain the
posterior and predictive distributions of $\theta_t|\Sigma=S,y^t$
and $y_{t+h}|\Sigma=S,y^t$, for a positive integer $h>0$, known as
the forecast horizon. Theorem \ref{th1} motivates approximating the
true posterior mean $S_t$ by $S=\widetilde{S}_t$, which is produced
from application of equation (\ref{eq3}), given a particular data
set $y^t=(y_1,y_2,\ldots,y_t)$. Thus we obtain the following
algorithm:

\begin{alg}\label{alg1}
\begin{enumerate}
\item [(a)] Prior distribution at time $t=0$:
$\theta_0|\Sigma=\widetilde{S}_0\sim
\mathcal{N}_d(\widetilde{m}_0,\widetilde{P}_0)$, for some
$\widetilde{m}_0$, $\widetilde{P}_0$ and $\widetilde{S}_0$. \item
[(b)] Posterior distribution at time $t$:
$\theta_{t}|\Sigma=\widetilde{S}_{t},y^{t}\sim
\mathcal{N}_d(\widetilde{m}_{t},\widetilde{P}_{t})$, where
$\widetilde{e}_t=y_t-\widetilde{y}_{t-1}(1)$ and
\begin{gather*}
\widetilde{m}_{t}=G\widetilde{m}_{t-1}+A_t\widetilde{e}_t,\quad
\widetilde{P}_{t}=G\widetilde{P}_{t-1}G'+\Omega-A_{t}\widetilde{Q}_tA_{t}',\quad
A_{t}=(G\widetilde{P}_{t-1}G'+\Omega)F\widetilde{Q}_t^{-1}, \\
\widetilde{S}_t=\frac{1}{n_0+t}\left(n_0\widetilde{S}_0+\sum_{i=1}^t\widetilde{S}_{i-1}^{1/2}\widetilde{Q}_i^{-1/2}
\widetilde{e}_i\widetilde{e}_i'\widetilde{Q}_i^{-1/2}\widetilde{S}_{i-1}^{1/2}\right).
\end{gather*}
\item [(c)] $h$-step forecast distribution at $t$:
$y_{t+h}|\Sigma=\widetilde{S}_t,y^t\sim
\mathcal{N}_p\{\widetilde{y}_t(h),\widetilde{Q}_t(h)\}$, where
$\widetilde{y}_t(h)=F'G^h\widetilde{m}_t$ and
$$
\widetilde{Q}_t(h)=F'G^h\widetilde{P}_t(G^h)'F +
\sum_{i=0}^{h-1}F'G^i\Omega (G^i)'F+\widetilde{S}_t.
$$
\end{enumerate}
\end{alg}
In the special case of matrix-variate DLMs (Harvey, 1986; West and
Harrison, 1997, \S 16.4) the estimator $S_t$ approximates the true
posterior mean of $\Sigma$ produced by an application of Bayes'
theorem, assuming a prior inverted Wishart distribution for
$\Sigma$. To see this, note that in the matrix-variate DLM (this
model is briefly in page \pageref{mvdlm}, see equation
(\ref{mvdlm})), $F$ is a $d$-dimensional design vector and
$Q_t=U_tS_{t-1}$ with $U_t=F'R_tF+1$ and so equation (\ref{eq3}) can
be written recursively as
\begin{gather}  \label{equnis1}
S_t = n_t^{-1}(n_{t-1}S_{t-1}+e_te_t'/U_t)\quad\textrm{and}\quad
n_t=n_{t-1}+1=n_0+t.
\end{gather}
It is easy to verify that the assumption
$\lim_{t\rightarrow\infty}S_t=\Sigma$ is satisfied, since
$\lim_{t\rightarrow\infty}S_t=\lim_{t\rightarrow\infty}\E(\Sigma|y^t)$
and
$\lim_{t\rightarrow\infty}\textrm{Var}\{\textrm{vech}(\Sigma)|y^t\}=0$,
where $\textrm{vech}(\cdot)$ denotes the column stacking operator of
a lower portion of a symmetric matrix. For $p=1$ the matrix-variate
DLM is reduced to the conjugate Gaussian/gamma DLM (West and
Harrison, 1997, \S 4.5). It turns out that the estimator $S_t$ of
equation (\ref{eq3}) approximates the analogous estimators of all
existing conjugate Gaussian dynamic linear models.

It is worth noting that Theorem \ref{th1} and Algorithm \ref{alg1}
have been presented for the state space model (\ref{model1}) having
time-invariant components $F$, $G$ and $\Omega$. However, these
results apply if some or all of the above components change with
time. In addition, if the evolution covariance matrix $\Omega_t$ is
time-dependent, it can be specified via discount factors (West and
Harrison, 1997, Chapter 6). This is a useful consideration, because
in practice the signal $\theta_t$ is unlikely to have the same
variability over time.

For the application of Algorithm \ref{alg1} the initial values
$\widetilde{m}_0$, $\widetilde{P}_0$, $n_0$ and $\widetilde{S}_0$
must be specified. $\widetilde{m}_0$ can be specified from
historical information from the underlying experiment and
$\widetilde{P}_0$ can be set as a typically large diagonal matrix,
e.g. $\widetilde{P}_0=1000 I_p$, reflecting a low precision (or
high uncertainty) on the specification of the moments of
$\theta_0$. The scalar $n_0$ can be set to $n_0=1$ (in the special
case of matrix-variate DLMs, $n_0$ is the prior degrees of
freedom). $\widetilde{S}_0$ is a prior estimate of $\Sigma$ and
requires at least a rough specification. As information is
deflated in time series, a miss-specification of $\widetilde{S}_0$
may not affect much the posterior estimate $\widetilde{S}_t$,
especially in the presence of large data sets. However, in many
cases and especially in financial time series, a
miss-specification of $\widetilde{S}_0$ can lead to poor estimates
of $\Sigma$. Here we suggest that a diagonal covariance matrix can
be used, where the diagonal elements of $\widetilde{S}_0$ reflect
the empirical expectation of the diagonal elements of $\Sigma$.
This expectation can be obtained by studying historical data and
other qualitative pieces of information, which are usually
available to practicing experts of the experiment or of the
application of interest.

\section*{Simulation Studies}

\subsection*{Empirical Convergence of
$\widetilde{S}_t$}

We have generated 1000 bivariate time series
$\{y_{it}\}^{t=1,2,\ldots,500}_{i=1,2,\ldots,1000}$ from several
state space models and then we have averaged the 1000 estimates
$\widetilde{S}_{i,t}$ (produced by each of the 1000 time series)
and compared the average
$\overline{\widetilde{S}}_t=1000^{-1}\sum_{i=1}^{1000}\widetilde{S}_{i,t}$
with the true value of $\Sigma$.

Since in practice complicated models are decomposed into simple
models comprising local level, polynomial trend and seasonal
components (Godolphin and Triantafyllopoulos, 2006), we consider
estimation separately in such different component models. We have
three modelling situations of interest: situation 1 (bivariate
local level models); situation 2 (bivariate linear trend models);
and situation 3 (bivariate seasonal models). For each of the above
three situations we have generated 1000 bivariate time series,
each of length 500, using three different covariance matrices
$\Sigma$, i.e.
$$
\Sigma_1 = \left[\begin{array}{cc} 2 & 3\\ 3 &
5\end{array}\right],\quad \Sigma_2 = \left[\begin{array}{cc} 100 &
85\\ 85 &
80\end{array}\right]\quad \textrm{and}\quad \Sigma_3=\left[\begin{array}{cc} 1 & 7\\
7 & 50\end{array}\right].
$$
Throughout the simulations we have chosen high correlations for each
$\Sigma_i$ $(i=1,2,3)$, since uncorrelated or approximately
uncorrelated state space models can be handled easily by employing
several univariate state space models. The priors of $\Sigma_i$ are
chosen as $\widetilde{S}_{i,0}^{(1)}=I_2$,
$\widetilde{S}_{i,0}^{(2)}=150I_2$ and
$\widetilde{S}_{i,0}^{(3)}=\textrm{diag}\{3,40\}$
$(i=1,2,\ldots,1000)$. The diagonal choice for the priors
$\widetilde{S}_{i,0}^{(j)}$ has been done for: (a) operational
simplicity (the user is likely to expect rough values for the
diagonal elements of $\Sigma_i$, rather than for the associated
correlations) and (b) judging how the estimation of $\Sigma_i$ is
affected by improper priors in the sense of setting the off-diagonal
elements of $\widetilde{S}_{i,0}^{(j)}$ to zero, while the true
values of $\Sigma_i$ posses high correlations. Throughout the models
the remaining settings are $n_0=1$, $\Omega=I_2$, $m_0=[0~0]'$ and
$P_0=1000I_2$, for all models. Table \ref{table1} shows the results.
There are three blocks of columns, each showing results of the state
space model considered, namely local level model (LL or block 1),
linear trend model (LT or block 2) and seasonal model (SE or block
3). In each block the first column shows the mean of the average
$\overline{\overline{S}}=500^{-1}\sum_{t=1}^{500}\overline{\widetilde{S}}_t$
of all $\overline{\widetilde{S}}_{t}$. The second column shows the
average $\overline{\widetilde{S}}_{100}$ at time point $t=100$.
Likewise the third column shows the respective
$\overline{\widetilde{S}}_{500}$ averaged over all 1000 series. The
rows in Table \ref{table1} show the picture of
$\overline{\widetilde{S}}_t$ over the three different values of
$\Sigma$, e.g. $\Sigma_1$, $\Sigma_2$ and $\Sigma_3$. The average
estimate of the correlations is also shown and it is marked in the
table by $\rho$. The results suggest that, generally, the LL model
has the best performance as opposed to the LT and the SE model,
although we note that $\sigma_{12}=3$ (covariance in $\Sigma_1$) is
estimated better from the LT model. It appears that the estimator
$\widetilde{S}_t$ for all models converges to the true values of
$\Sigma$, but the rate of convergence depends on the underlying
state space model (here LL performs faster convergence) and on the
prior $\widetilde{S}_0$.

Table \ref{table2} shows the averaged (over all 1000 simulated time
series) mean vector of squared standardized one-step forecast errors
$(\textrm{MSSE}^{(1)})$, for each of the three models (LL, LT, SE)
and for each of $\Sigma$ $(\Sigma_1,\Sigma_2,\Sigma_3)$. For
comparison purposes, Table \ref{table2} also shows the respective
values of the $\textrm{MSSE}^{(2)}$ when $\Sigma_i$ is the true
value. The target value of the $\textrm{MSSE}^{(i)}$ is $[1~1]$. We
see that the $\textrm{MSSE}^{(1)}$ approaches the respective
$\textrm{MSSE}^{(2)}$ and this demonstrates the accuracy of the
estimator $\widetilde{S}_t$. We observe that under $\Sigma_3$, the
$\textrm{MSSE}^{(1)}$ has values significantly smaller than 1 as
compared to the $\textrm{MSSE}^{(2)}$ using the true value of
$\Sigma_3$.

\begin{table}
\caption{Performance of the estimator $\widetilde{S}_t$ (of
Algorithm \ref{alg1}) for 1000 simulated bivariate dynamic models
generated from a local level model (LL), a linear trend model (LT)
and a seasonal model (SE) under three observation covariance
matrices $\Sigma_1$, $\Sigma_2$ and $\Sigma_3$.}\label{table1}
\begin{center}
\begin{tabular}{|c||ccc||ccc||ccc|}
\hline Model & LL &  & & LT &  & & SE &  & \\
$\Sigma=\Sigma_i$ & $\overline{\overline{S}}$ &
$\overline{\widetilde{S}}_{100}$ & $\overline{\widetilde{S}}_{500}$
& $\overline{\overline{S}}$ & $\overline{\widetilde{S}}_{100}$ &
$\overline{\widetilde{S}}_{500}$ & $\overline{\overline{S}}$ & $\overline{\widetilde{S}}_{100}$ & $\overline{\widetilde{S}}_{500}$ \\
\hline $\sigma_{11}=2$ &
1.945 & 1.938 & 1.997 & 2.572 & 2.721 & 2.392 & 2.171 & 2.207 & 2.165 \\
$\sigma_{12}=3$ & 2.798 & 2.770 & 2.920 & 2.988 & 3.029 & 3.026 & 2.314 & 2.186 & 2.589 \\
$\sigma_{22}=5$ & 4.722 & 4.685 & 4.899 & 4.547 & 4.489 & 4.777 & 4.399 & 4.283 & 4.694 \\
$\rho=0.948$ & 0.923 & 0.919 & 0.933 & 0.874 & 0.867 & 0.895 &
0.748 & 0.711 & 0.812 \\ \hline
 $\sigma_{11}=100$ & 100.039 & 99.931 & 100.303 & 98.271 & 98.157 & 98.368 & 98.731 & 98.806 & 99.602 \\
$\sigma_{12}=85$ & 83.133 & 83.028
& 84.757 & 79.471 & 78.969 & 82.660 & 79.627 & 79.427 & 83.254  \\
$\sigma_{22}=80$ & 80.430 & 80.277 & 80.353 & 78.917 & 78.865 & 79.822 & 79.755 & 79.886 & 79.896  \\
$\rho=0.950$ & 0.927 & 0.927 & 0.944 & 0.902 & 0.897 & 0.933 &
0.897 & 0.894 & 0.933 \\ \hline
 $\sigma_{11}=1$ & 1.124 & 1.135 & 1.101 & 1.200 & 1.234 & 1.126 & 1.184 & 1.202 & 1.151 \\
$\sigma_{12}=7$ & 6.506 & 6.457 & 6.735 & 5.388 & 5.177 & 5.904 & 5.764 & 5.623 & 6.234  \\
$\sigma_{22}=50$ & 49.305 & 49.375 & 49.840 & 48.518 & 48.579 & 49.392 & 48.816 & 49.038 & 49.540 \\
$\rho=0.989$ & 0.784 & 0.862 & 0.909 & 0.706 & 0.668 & 0.791 & 0.758 & 0.732 & 0.825 \\
\hline
\end{tabular}
\end{center}
\end{table}

\begin{table}
\caption{Mean vector of squared standardized one-step forecast
errors (MSSE$^{(i)}$) of the multivariate dynamic model of Algorithm
\ref{alg1}. The index $i=1,2$ refers to when $\Sigma$ is estimated
by the data $(i=1)$ and when $\Sigma$ is assumed known (according to
the simulations) for comparison purposes $(i=2)$. The notation LL,
LT, SE and $\Sigma_1$, $\Sigma_2$, $\Sigma_3$ is the same as in
Table \ref{table1}.}\label{table2}
\begin{center}
\begin{tabular}{|c||cc||cc|}
\hline & MSSE$^{(1)}$ &  & MSSE$^{(2)}$ &    \\
\hline LL $(\Sigma_1)$ & 0.994 & 1.071  & 0.999 & 0.995 \\
LL $(\Sigma_2)$ & 0.939 & 0.914  & 0.999 & 0.999 \\
LL $(\Sigma_3)$ & 0.773 & 1.026  & 0.998 & 0.997 \\
LT $(\Sigma_1)$ & 0.875 & 1.141  & 0.992 & 0.996 \\
LT $(\Sigma_2)$ & 0.900 & 0.895  & 1.002 & 0.997 \\
LT $(\Sigma_3)$ & 0.774 & 1.031  & 1.000 & 0.996 \\
SE $(\Sigma_1)$ & 0.930 & 1.092  & 0.997 & 0.999 \\
SE
$(\Sigma_2)$ & 0.903 & 0.864 & 0.998 & 0.996 \\
SE $(\Sigma_3)$ & 0.805 & 1.026 & 0.998 & 0.996 \\
\hline
\end{tabular}
\end{center}
\end{table}

\subsection*{Comparison of the Local Level Model with
MCMC}

We have simulated a single local level model under the observation
covariance matrix $\Sigma=\Sigma_1$ and the relevant model
components of the local level model of the previous sub-section. We
apply Algorithm \ref{alg1} and we compare it with a state of the art
MCMC estimation procedure based on a blocked Gibbs sampler suitable
for state space models (Gamerman, 1997, p. 149); the MCMC procedure
we use is described in the appendix. The MCMC estimation procedure
is an iterative non-sequential MCMC procedure and its role in this
section is to provide a means of comparison with the non-iterative
procedure of Algorithm \ref{alg1}. MCMC is the \emph{gold standard},
since it produces (given enough computation) exact computation of
$S_t$. But MCMC is impractical; the new proposed method is a quick,
practical and easily implemented approximation. In this section we
compare the new method with the gold standard in order to show how
good is the approximation. Tables \ref{table3} and \ref{table4} give
the results; the former shows the estimates of $\Sigma$ with both
methods (MCMC and Algorithm \ref{alg1}) and the latter shows the
performance of the one-step forecast errors for both methods. In
Table \ref{table4} the one-step forecast error vector
$e_t=[e_{1t}~e_{2t}]'$ and the mean vector of squared one-step
forecast errors are shown for several values of $t$ under both
estimation methods. We observe that the new method (of Algorithm
\ref{alg1}) approximates well the MCMC estimates, especially for
large values of time $t=N$.

We note that MCMC should not be considered as a better method as
compared to the proposal of Algorithm \ref{alg1}, since MCMC is an
iterative and in particular in this paper it is a non-sequential
estimation procedure. The application of sequential MCMC estimation
(Doucet {\it et. al.}, 2001) often experience several challenges as
for example time-constraints, availability for general purpose
algorithms, prior-specification, prior-sensitivity, fast monitoring
and expert intervention features. The proposal of this paper
provides a strong modelling approach allowing for variance
estimation in a wide class of conditionally Gaussian dynamic linear
models and this section shows that for large time periods its
performance is close to Monte Carlo estimation.

\begin{table}
\caption{Bivariate simulated local level dynamic linear model.
Showed are: real versus estimated values of
$\Sigma=\{\sigma_{ij}\}_{i,j=1,2}$ and the correlation coefficient
$\rho$. The first column indicates how many observations were used
in each estimation.} \label{table3}
\begin{center}
\begin{tabular}{|c||c|c|c||c|c|c||c|c|c||c|c|c|}
\hline & \multicolumn{3}{|c||}{$\sigma_{11}$} &
\multicolumn{3}{|c||}{$\sigma_{12}$} &
\multicolumn{3}{|c||}{$\sigma_{22}$} &
\multicolumn{3}{|c|}{$\rho$} \\
\multicolumn{1}{|c||}{$N$} & \multicolumn{1}{c}{Real} &
\multicolumn{1}{c}{MCMC} &\multicolumn{1}{c||}{New} &
\multicolumn{1}{c}{Real} & \multicolumn{1}{c}{MCMC} &
\multicolumn{1}{c||}{New} & \multicolumn{1}{c}{Real} &
\multicolumn{1}{c}{MCMC} & \multicolumn{1}{c||}{New} &
\multicolumn{1}{c}{Real} & \multicolumn{1}{c}{MCMC} & \multicolumn{1}{c|}{New}  \\
\hline
100 & 2.00 & 1.68 & 1.24 & 3.00 & 2.35 & 1.82 & 5.00 & 4.12 & 3.69 & 0.95 & 0.90 & 0.85 \\
150 & 2.00 & 1.78 & 1.34 & 3.00 & 2.44 & 1.91 & 5.00 & 3.97 & 3.72 & 0.95 & 0.92 & 0.86 \\
200 & 2.00 & 1.76 & 1.46 & 3.00 & 2.48 & 2.04 & 5.00 & 4.02 & 3.77 & 0.95 & 0.94 & 0.87 \\
250 & 2.00 & 2.04 & 1.64 & 3.00 & 2.89 & 2.33 & 5.00 & 4.50 & 4.17 & 0.95 & 0.95 & 0.89 \\
300 & 2.00 & 1.92 & 1.65 & 3.00 & 2.78 & 2.39 & 5.00 & 4.44 & 4.31 & 0.95 & 0.95 & 0.90 \\
350 & 2.00 & 1.90 & 1.69 & 3.00 & 2.76 & 2.46 & 5.00 & 4.46 & 4.40 & 0.95 & 0.95 & 0.90 \\
400 & 2.00 & 2.01 & 1.76 & 3.00 & 2.90 & 2.56 & 5.00 & 4.59 & 4.50 & 0.95 & 0.96 & 0.91 \\
450 & 2.00 & 2.05 & 1.82 & 3.00 & 2.97 & 2.66 & 5.00 & 4.71 & 4.65 & 0.95 & 0.96 & 0.92 \\
500 & 2.00 & 2.11 & 1.85 & 3.00 & 3.09 & 2.74 & 5.00 & 4.90 & 4.79 & 0.95 & 0.96 & 0.92 \\
\hline
\end{tabular}
\end{center}
\end{table}

\begin{table}
\caption{Bivariate simulated local level dynamic linear model.
Showed are: one-step forecast errors at time $t=N$ and the squared
sums of the forecasting errors up to time $N$.} \label{table4}
\begin{center}
\begin{tabular}{|c||c|c||c|c||c|c||c|c|}
\hline & \multicolumn{2}{|c||}{$e_{1N}$} &
\multicolumn{2}{|c||}{$e_{2N}$} &
\multicolumn{2}{|c||}{$N^{-1}\sum_t^N e_{1t}^2$} &
\multicolumn{2}{|c|}{$N^{-1}\sum_t^N e_{2t}^2$} \\
\multicolumn{1}{|c||}{$N$}  & \multicolumn{1}{c}{MCMC} &
\multicolumn{1}{c||}{New} & \multicolumn{1}{c}{MCMC} &
\multicolumn{1}{c||}{New} & \multicolumn{1}{c}{MCMC} &
\multicolumn{1}{c||}{New} &
\multicolumn{1}{c}{MCMC} & \multicolumn{1}{c|}{New}  \\
\hline
100 & $-$2.14  & $-$2.22 & $-$2.46  & $-$2.49 & 4.85  & 4.84 & 8.54  & 8.59 \\
150 & $-$0.14  & $-$0.28 & $-$3.88  & $-$3.92 & 4.83  & 4.86 & 8.21  & 8.26 \\
200 & $-$1.02  & $-$1.09 & $-$0.12  & 0.05    & 5.02  & 5.04 & 8.10  & 8.13 \\
250 & $-$0.24  & $-$0.20 & $-$1.45  & $-$1.47 & 5.25  & 5.30 & 8.72  & 8.76 \\
300 & $-$1.72  & $-$1.79 & $-$0.61  & $-$0.71 & 5.01  & 5.12 & 8.91  & 8.95 \\
350 & $-$0.42  & $-$0.46 & $-$1.91  & $-$1.91 & 5.12  & 5.12 & 9.01  & 9.04 \\
400 & $-$0.42  & $-$0.54 & $-$2.26  & $-$2.29 & 5.21  & 5.22 & 9.09  & 9.12 \\
450 & $-$3.91  & $-$4.02 & $-$5.42  & $-$5.41 & 5.28  & 5.29 & 9.32  & 9.33 \\
500 & 1.48     & 1.68    & $-$0.93  & $-$0.98 & 5.24  & 5.24 & 9.54  & 9.54 \\
\hline
\end{tabular}
\end{center}
\end{table}

\section*{Application to VAR and TVVAR Time Series
Models}

The dynamic model (\ref{model1}) is very general and an important
subclass of (\ref{model1}) is the popular vector ARMA model. In
recent years vector autoregressive (VAR) models have been
extensively developed and used, especially for economic time
series, as in Doan {\it et al.} (1984), Litterman (1986), Kadiyala
and Karlsson (1993, 1997), Ooms (1994), Johansen (1995), Uhlig
(1997), Ni and Sun (2003), Sun and Ni (2004) and Huerta and Prado
(2006).

Our discussion in this section includes two important subclasses
of model (\ref{model1}), which can be used for a wide-class of
stationary and non-stationary time series forecasting. The first
is the VAR model of known order $\ell\geq 1$, defined by
\begin{equation}\label{var1}
y_t=\Phi_1y_{t-1}+\Phi_2y_{t-2}+\cdots+\Phi_\ell
y_{t-\ell}+\epsilon_t,\quad \epsilon_t \sim \mathcal{N}_p(0,\Sigma),
\end{equation}
where $\Phi_1,\Phi_2,\ldots,\Phi_\ell$ are $p\times p$ matrices of
parameters. In the usual estimation of VAR, stationarity has to be
assumed and so the roots of the polynomial (in $z$)
$$
|I_p-\Phi_1 z - \Phi_2 z^2-\cdots-\Phi_\ell z^\ell|=0
$$
should lie outside the unit circle. In standard theory (\ref{var1})
may not assume a Gaussian distribution for $\epsilon_t$, although in
practice this is used for operational simplicity. It is also known
that for a high order $\ell$ model (\ref{var1}) approximates
multivariate moving average models, which are typically difficult to
estimate and this makes the VAR even more attractive in
applications. It is also known that for general on-line estimation
and forecasting, the covariance matrix $\Sigma$ either has to be
assumed known or it has to be diagonal. This is a major limitation,
because it means that either the modeller knows \emph{a priori} the
cross-correlation between the series
$\{y_{1t}\},\{y_{2t}\},\ldots,\{y_{pt}\}$, where
$y_t=[y_{1t}~y_{2t}~\cdots~y_{pt}]'$, or that the $p$ scalar time
series are all stochastically uncorrelated, in which case it is more
sensible to use several univariate AR models instead. Recently, the
need for estimation of $\Sigma$ as a full covariance matrix (e.g.
where $\Sigma$ has $p(p+1)/2$ elements to be estimated) is
considered, but the existing estimation procedures include
necessarily iterative estimation via importance sampling (Kadiyala
and Karlsson, 1997). Ni and Sun (2003) point out that from a
frequentist standpoint ordinary least squares and maximum likelihood
estimators of (\ref{var1}) are unavailable. These authors state that
asymptotic theory estimators may not be applicable for VAR
(especially when $\{y_t\}$ is a short-length time series). Ni and
Sun (2003), Sun and Ni (2004) and Huerta and Prado (2006) propose
Bayesian estimation of the autoregressive parameters $\Phi_i$ and
$\Sigma$, based on MCMC. It follows that for model (\ref{model1})
when $\Sigma$ is unknown, only iterative estimation procedures can
be applied. Our proposal for on-line estimation of $\Sigma$ gives a
step forward to the estimation and forecasting of VAR models and it
is outlined below.

We propose a generalization of the univariate state space
representation considered in West and Harrison (1997, \S 9.4.6).
Other state space representations of the VAR are considered in
Huerta and Prado (2006), but these representations, usually
referred to as canonical representations of the VAR model (Shumway
and Stoffer, 2000) are not convenient for the estimation of
$\Sigma$, because $\Sigma$ is embedded into the evolution equation
of the states $\theta_t$. First note that we can rewrite
(\ref{var1}) as $y_t=\Phi X_t+\epsilon_t$, where $\Phi =
[\Phi_1~\Phi_2~\cdots~\Phi_\ell ]$ and
$X_t=[y_{t-1}'~y_{t-2}'~\cdots~y_{t-\ell}']'$ and so we can write
\begin{equation}\label{rep2}
y_t=F_t'\theta + \epsilon_t = (X_t'\otimes
I_p)\textrm{vec}(\Phi)+\epsilon_t,
\end{equation}
where $\textrm{vec}(\cdot)$ denotes the column stacking operator
of a portion of a matrix and $\otimes$ denotes the Kronecker or
tensor product of two matrices. Model (\ref{rep2}) can be seen as
a regression-type time series model and it can be handled by the
general Algorithm \ref{alg1} for model (\ref{model1}) if we set
$G=I_p$, $\Omega=0$ and if we replace $F$ by the time-varying
$F_t=X_t\otimes I_p$. Thus we can readily apply Algorithm
\ref{alg1} to estimate $\Sigma$ and $\theta$ or
$\Phi_1,\Phi_2,\ldots,\Phi_{\ell}$.

Moving to the time-varying vector autoregressive (TVVAR) time
series, in recent years there has been a growing literature for
TVVAR time series. Kitagawa and Gersch (1996), Dahlhaus (1997),
Francq and Gautier (2004) and Anderson and Meerschaert (2005) study
parameter estimation based on the asymptotic behaviour of TVVAR and
time-varying ARMA models. From a state space standpoint West {\it et
al.} (1999) propose a state space formulation for a univariate
time-varying AR model applied to electroencephalographic data. In
this section we extend this state space formulation to a vector of
observations and hence we can propose the application of Algorithm
\ref{alg1} in order to estimate the covariance matrix of the error
drifts of the TVVAR model.

Consider that the $p$-vector time series $\{y_t\}$ follows the
TVVAR model of known order $\ell$ defined by
\begin{equation}\label{var2}
y_t=\Phi_{1t}y_{t-1}+\Phi_{2t}y_{t-2}+\cdots+\Phi_{\ell t}
y_{t-\ell}+\epsilon_t,\quad \epsilon_t \sim \mathcal{N}_p(0,\Sigma),
\end{equation}
where $\Phi_{1t},\Phi_{2t},\ldots,\Phi_{\ell t}$ are the
time-varying autoregressive parameter matrices. The model can be
stationary, locally-stationary or non-stationary depending on the
roots of the $t$ polynomials (in $z$)
$$
|I_p-\Phi_{1t} z - \Phi_{2t} z^2-\cdots-\Phi_{\ell t} z^\ell|=0.
$$
Typical considerations include the local stationarity where there
are several regimes for which, locally,  $\{y_t\}$ is stationary,
but globally $\{y_t\}$ is non-stationary. Also the time-dependent
parameter matrices $\Phi_{it}$ can allow for an improved dynamic fit
as opposed to the static parameters of the VAR.

In our development we adopt a random walk for the evolution of the
parameters $\Phi_{it}$ $(i=1,2,\ldots,\ell)$, although the
modeller might suggest other Markovian stochastic evolution
formulae for $\Phi_{it}$. The random walk evolution is the natural
consideration when $\{y_t\}$ is assumed locally stationary. Hence
we can rewrite model (\ref{var2}) in state-space form as
\begin{equation}\label{rep3}
y_t=\Phi_tX_t+\epsilon_t=F_t'\theta_t+\epsilon_t\quad\textrm{and}\quad
\theta_t=\theta_{t-1}+\omega_t,
\end{equation}
where $X_t=[y_{t-1}'~y_{t-2}'~\cdots~y_{t-\ell}']'$, $F_t=X_t\otimes
I_p$, $\Phi_t = [\Phi_{1t}~\Phi_{2t}~\cdots~\Phi_{\ell t}]$,
$\theta_t=\textrm{vec}(\Phi_t)$ and $\omega_t\sim
\mathcal{N}_{p^2\ell}(0,\Omega)$, for some transition covariance
matrix $\Omega$. Model (\ref{rep3}) is reduced to (\ref{rep2}) when
$\Omega=0$, in which case $\theta_t=\theta_{t-1}=\theta$. After
specifying $\Omega$, we can directly apply Algorithm \ref{alg1} to
the state space model (\ref{rep3}) and thus we can obtain an
algorithm for the estimation of $\Sigma$, for the estimation of
$\theta_t$ or $\Phi_{1t},\Phi_{2t},\ldots,\Phi_{\ell t}$ and for
forecasting the series $\{y_t\}$.

\begin{figure}
 \epsfig{file=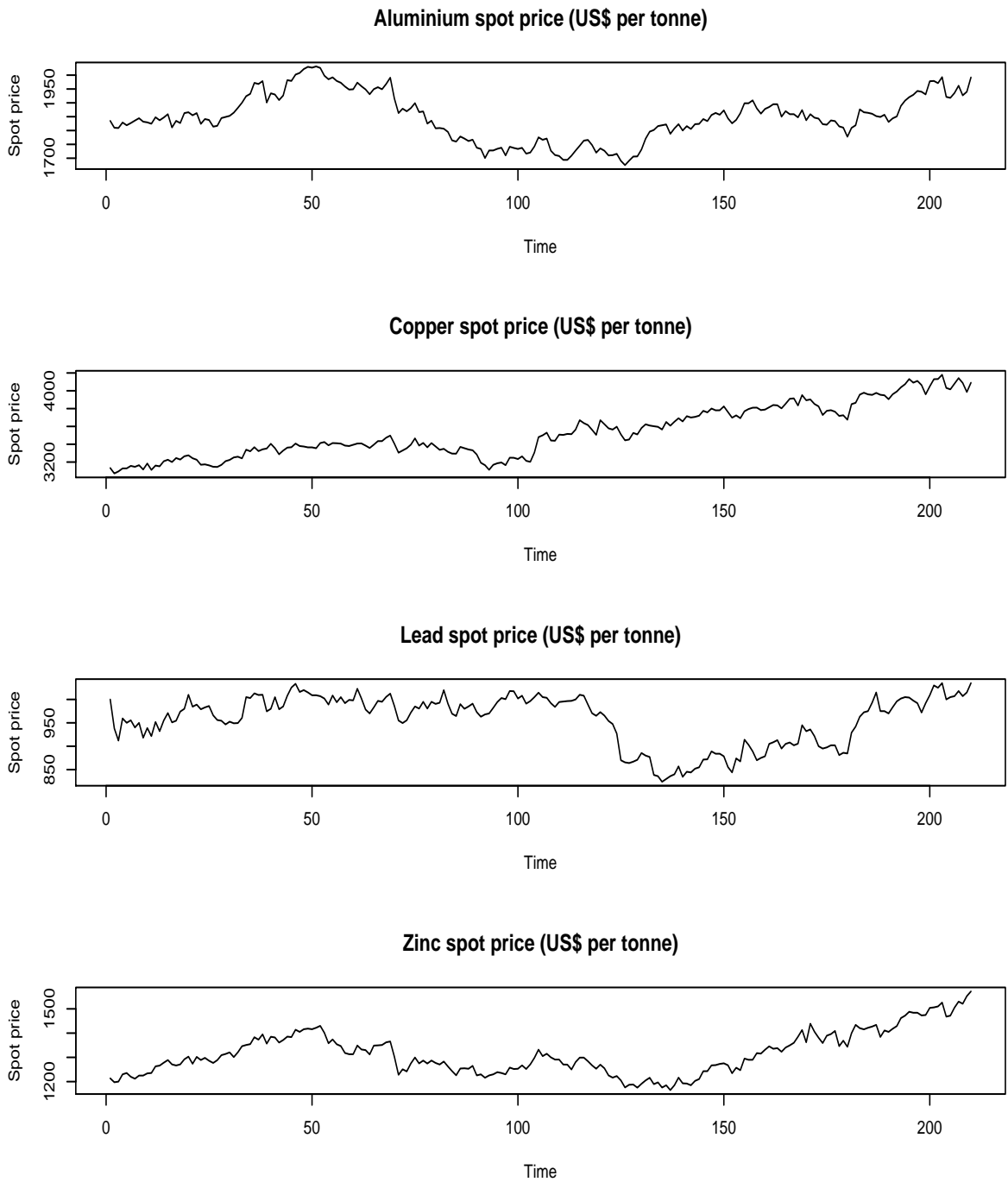, height=19cm, width=16cm}
 \caption{LME data, consisting of aluminium, copper, lead and zinc spot prices (in US dollars
 per tonne of each metal).}\label{fig1}
\end{figure}

\section*{London Metal Exchange Data}

In this section we analyze London metal exchange (LME) data
consisting of official spot prices (US dollars per tonne of metal).
LME is the world's leading non-ferrous metals' market, trading
currently highly liquid contracts for metals, such as aluminium,
aluminium alloy, copper, lead, nickel, tin and zinc. According to
the LME website ({\tt http://www.lme.co.uk/}) ``LME is highly
successful with a turnover in excess of US\$3,000 billion per annum.
It also contributes to the UK's invisible earnings to the sum of
more than \pounds 250 million in overseas earnings each year." More
information about the functions of the LME can be found via its
website (see above); the recently growing literature on the
econometrics modelling of the LME can be found in the review of
Watkins and McAleer (2004).

We consider forecasting for four metals exchanged in the LME,
namely aluminium, copper, lead and zinc. The data are provided
from the LME website for the period of 4 January 2005 to 31
October 2005. After excluding weekends and bank holidays there are
$N=210$ trading days. We store the data into the $4\times 1$
vector time series $\{y_t\}_{t=1,2,\ldots,210}$ and
$y_t=[y_{1t}~y_{2t}~y_{3t}~y_{4t}]'$, where $y_{1t}$ denotes the
spot price at time $t$ of aluminium, $y_{2t}$ denotes the spot
price at time $t$ of copper, $y_{3t}$ denotes the spot price at
time $t$ of lead and $y_{4t}$ denotes the spot price at time $t$
of zinc. The data are plotted in Figure \ref{fig1}.

We propose the VAR and TVVAR models of the previous section; the
motivation of this being that from Figure \ref{fig1} the evolution
of the data seems to follow roughly an autoregressive type model.
Indeed there is an apparent trend with no seasonality, which can
be modelled with a trend model or with a VAR or TVVAR model of the
previous section. Here we illustrate the proposal of VAR and TVVAR
models, which, according to the previous section, can estimate the
covariance matrix of $y_t$, given the state parameters, and thus
the correlation structure of $\{y_t\}$ can be studied. Other
models for this kind of data have been applied in
Triantafyllopoulos (2006) and we can envisage that the models of
West and Quintana (1987) can also be applied to the LME data.

First we apply the algorithms of the previous section to several VAR
and TVVAR models of different orders in order to find out which
model gives the best performance. Performance here is measured via
the mean vector of squared standardized one-step forecast errors
(MSSE) and the mean vector of absolute percentage one-step forecast
errors (MAPE). The first is chosen as a general performance measure
taking into account the estimation of the covariance matrix $\Sigma$
and the second is chosen as a generally reliable percent performance
measure. Table \ref{table7} shows the results of 10 VAR$(i)$ and
TVVAR$(i)$ models (first column) of order $i=1,2,\ldots,10$. The
discount factor $\delta$ refers to the discounting of the evolution
covariance matrix of the state parameters $\theta_t$; $\delta=1$
refers to a static $\theta_t=\theta$ (VAR model), while $\delta<1$
refers to a dynamic local level evolution of
$\theta_t=\theta_{t-1}+\omega_t$ (TVVAR model). Table \ref{table7}
shows that the performance of the TVVAR is remarkable compared with
the performance of VAR, which produces very high MSSE throughout the
range of $i$. Out of the VAR models, the best is the VAR(1), which
still produces very large MSSE. This indicates that a moving average
(MA) model is unlikely to produce good results at all, as the MSSE
of the VAR increases with the order $i$. Also the approximation of a
MA model with a high order VAR model will include a large number of
state parameters to be estimated and this will introduce
computational problems.

\begin{table}
\caption{Mean vector of squared standardized one-step forecast
errors (MSSE) and mean vector of absolute percentage one-step
forecast errors (MAPE) of the multivariate LME time series
$\{y_t\}$. The first column indicates several VAR and TVVAR
models.}\label{table7}
\begin{center}
\begin{tabular}{|c||cccc||cccc|}
\hline & MSSE & & & & MAPE & & & \\ \hline VAR(1) & 6.614 &
16.782 & 7.655 & 18.370 & 0.033 & 0.071 & 0.143 & 0.057 \\
TVVAR(1): $\delta=0.1$ & 2.430 & 1.764 & 0.622 & 1.852 & 0.059 &
0.053 & 0.084 & 0.076 \\ VAR(2) & 19.610 & 15.966 & 11.934 &
10.271 & 0.081 & 0.226 & 0.201 & 0.101 \\ TVVAR(2): $\delta=0.35$
& 1.296 & 1.743 & 1.228 & 1.822 & 0.065 & 0.057 & 0.116 & 0.095 \\
VAR(3) & 11.777 & 23.715 & 9.906 & 9.058 & 0.585 & 0.345 & 0.480 &
0.246 \\ TVVAR(3): $\delta=0.65$ & 2.254 & 3.149 & 2.222 & 2.180 &
0.074 & 0.053 & 0.132 & 0.108 \\ VAR(4) & 39.979 & 54.892 & 28.407
& 19.169 & 0.159 & 0.103 & 0.235 & 0.161 \\ TVVAR(4): $\delta=0.6$
& 1.389 & 1.802 & 1.210 & 1.329 & 0.101 & 0.072 & 0.179 & 0.147 \\
VAR(5) & 18.592 & 16.605 & 15.474 & 12.570 & 0.203 & 0.076 & 0.392
& 0.248 \\ TVVAR(5): $\delta=0.7$ & 1.429 & 2.269 & 1.651 & 1.677
& 0.114 & 0.079 & 0.208 & 0.171 \\ VAR(6) & 24.910 & 19.085 &
14.584 & 17.784 & 0.206 & 0.134 & 0.320 & 0.197 \\ TVVAR(6):
$\delta=0.75$ & 1.828 & 2.705 & 1.683 & 1.757 & 0.132 & 0.089 &
0.243 & 0.197 \\ VAR(7) & 21.722 & 38.054 & 14.597 & 14.180 &
0.330 & 0.092 & 0.422 & 0.490 \\ TVVAR(7): $\delta=0.75$ & 1.366 &
2.044 & 1.191 & 1.531 & 0.148 & 0.101 & 0.280 & 0.223 \\ VAR(8) &
28.985 & 35.867 & 11.291 & 16.370 & 0.515 & 0.325 & 0.812 & 0.563
\\ TVVAR(8): $\delta=0.8$ & 2.130 & 2.900 & 1.637 & 1.910 & 0.168
& 0.111 & 0.326 & 0.249 \\ VAR(9) & 40.229 & 53.798 & 12.249 &
19.691 & 0.393 & 0.184 & 0.416 & 0.411 \\ TVVAR(9): $\delta=0.95$
& 14.042 & 21.011 & 6.724 & 8.708 & 0.207 & 0.124 & 0.352 & 0.284
\\ VAR(10) & 46.791 & 49.869 & 16.240 & 23.974 & 0.611 & 0.306 &
0.751 & 0.694 \\ TVVAR(10): $\delta=0.9$ & 4.273 & 7.541 & 3.637 &
5.629 & 0.205 & 0.124 & 0.391 & 0.296 \\
\hline
\end{tabular}
\end{center}
\end{table}

\begin{figure}[h]
 \epsfig{file=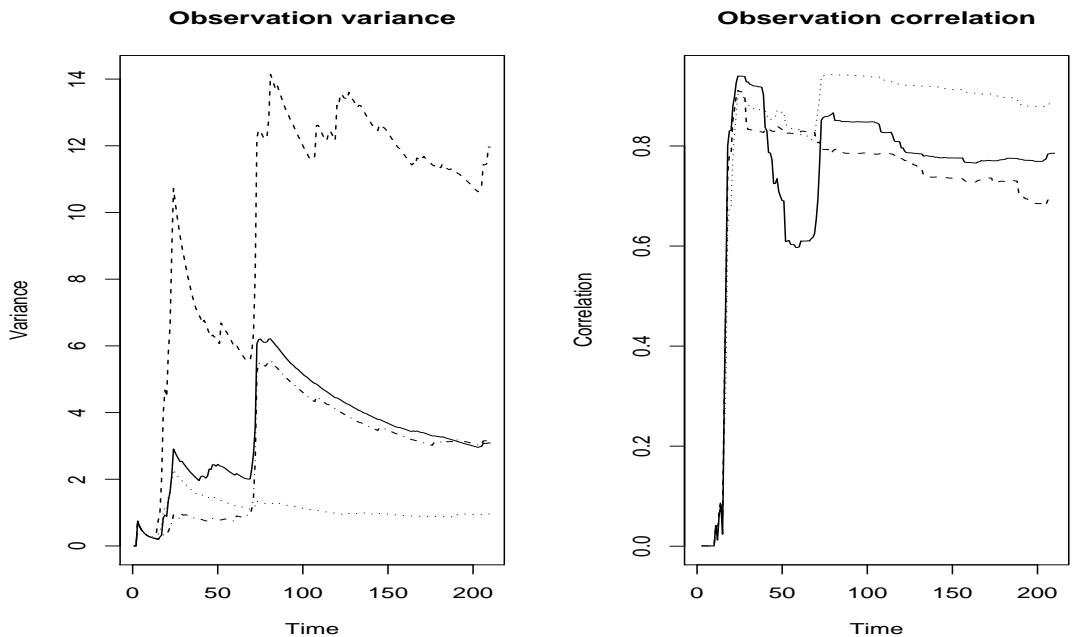, height=9cm, width=15cm}
 \caption{Estimate of the observation covariance matrix $\Sigma=\{\sigma_{ij}\}_{i,j=1,2,3,4}$. The left graph
 shows the estimates of the variances $\sigma_{ii}$; the solid line shows the estimate of $\sigma_{11}$,
 the dashed line shows the estimate of $\sigma_{22}$, the dotted line shows the estimate of $\sigma_{33}$,
 the dashed/dotted line shows the estimate of $\sigma_{44}$. The right graph shows the estimate of the
 correlations of $y_{1t}|\theta_t$ with $y_{jt}|\theta_t$ $(j=2,3,4)$; the solid line shows the
 estimate of the correlation of $y_{1t}|\theta_t$ with $y_{2t}|\theta_t$, the dashed line shows the estimate
 of the correlation of $y_{1t}|\theta_t$ with $y_{3t}|\theta_t$ and the dotted line shows the estimate
 of the correlation of $y_{1t}|\theta_t$ with $y_{4t}|\theta_t$.}\label{fig4}
\end{figure}

Therefore, our attention is focused on the TVVAR models. From a
computational standpoint we note that as the order increases
$\delta$ can not be too low, because then there are computational
difficulties in the calculation of the symmetric square root of
$\widetilde{Q}_t$, used for the estimation of $\widetilde{S}_t$ (the
estimate of $\Sigma$). Lower values of $\delta$ work better
(Triantafyllopoulos, 2006) and here we have chosen the lowest values
of $\delta$, which are allowed. Our decision on the best TVVAR model
is based on the following four criteria.
\begin{enumerate}
\item low order models are preferable as they have fewer state
parameters; \item $\delta$ should not be too low, because then the
covariance matrix of $\theta_t$ will be too large; \item the MSSE
vector should be close to $[1~1~1~1]'$; \item the MAPE vector should
be as low as possible.
\end{enumerate}
Considering the above criteria we favor the TVVAR(2). Figure
\ref{fig4} shows the estimate of the observation covariance matrix
$\Sigma$. From the right graph we observe that the estimate of the
correlations of $y_{1t}$ and $y_{jt}$, given $\theta_t$ are very
high (close to 1) and this means that in forecasting; this provides
useful information about the cross-dependence of the four metal
prices over time.

As mentioned before two competitive models to our TVVAR modelling
for the LME data are the matrix-variate DLMs (MV-DLMs) of Quintana
and West (1987) and the discount weighted regression (DWR) of
Triantafyllopoulos (2006). Next we compare the TVVAR(2) model
discussed above with these two modelling approaches. We start by
briefly describing the MV-DLM and the DWR.

The MV-DLM is defined by
\begin{equation}\label{mvdlm}
y_t'=F'\Theta_t+\epsilon_t'\quad \textrm{and}\quad
\Theta_t=G\Theta_{t-1}+\omega_t,
\end{equation}
where $F$ is a $d\times 1$ design vector, $\Theta_t$ is a $d\times
p$ state matrix, $G$ is a $d\times d$ transition matrix,
$\epsilon_t|\Sigma\sim \mathcal{N}_p(0,\Sigma)$ and
$\textrm{vec}(\omega_t)|\Sigma,\Omega \sim
\mathcal{N}_{dp}(0,\Sigma\otimes \Omega)$, where
$\textrm{vec}(\cdot)$ denotes the column stacking operator of a
lower portion of a matrix and $\otimes$ denotes the Kronecker
product of two matrices. A prior inverted Wishart distribution is
assumed for $\Sigma$ and the resulting posterior distributions as
well as further details on the model can be found in Quintana and
West (1987) and West and Harrison (1997, Chapter 16) (for more
references on this model, see also the Introduction). In the
application of MV-DLMs it is necessary to specify $F$ and $G$.
Following Quintana and West (1987), who consider international
exchange rates data, and by consulting the plots of Figure
\ref{fig1} we propose a linear trend model for the LME data. Thus we
can set
$$
F=\left[\begin{array}{c} 1\\
0\end{array}\right]\quad\textrm{and}\quad G=\left[\begin{array}{cc}
1 & 1 \\ 0 & 1\end{array}\right].
$$

\begin{table}
\caption{Mean vector of squared standardized one-step forecast
errors (MSSE) and mean vector of absolute percentage one-step
forecast errors (MAPE) for the LME data and for three multivariate
models: TVVAR, MV-DLM and DWR.}\label{table8}
\begin{center}
\begin{tabular}{|c||cccc||cccc|}
\hline & MSSE & & & & MAPE & & & \\ \hline TVVAR(2) & 1.296 & 1.743
& 1.228 & 1.822 & 0.065 & 0.057 & 0.116 & 0.095 \\ MV-DLM & 1.306 &
2.436 & 0.984 & 1.887 &
0.019 & 0.022 & 0.026 & 0.025 \\ DWR & 2.202 & 1.610 & 1.590 & 1.868 & 0.013 & 0.015 & 0.017 & 0.017 \\
\hline
\end{tabular}
\end{center}
\end{table}

The DWR is defined by
$$
y_t=y_{t-1}+\psi_t+\epsilon_t\quad \textrm{and}\quad
\psi_t=\psi_{t-1}+\zeta_t,
$$
with $\epsilon_t|\Sigma \sim \mathcal{N}_p(0,\Sigma)$ and
$\zeta_t\sim \mathcal{N}_p(0,\Omega_t)$. This model can be put into
state space form as in
$$
y_t=[y_{t-1}~I_p]\left[\begin{array}{c} 1 \\
\psi_t\end{array}\right] + \epsilon_t=F_t'\theta_t+\epsilon_t,\quad
\theta_t=\left[\begin{array}{c} 1
\\ \psi_t \end{array}\right] = \left[\begin{array}{c} 1
\\ \psi_{t-1} \end{array}\right] + \left[\begin{array}{c} 0
\\ \zeta_t \end{array}\right] = \theta_{t-1}+\omega_t.
$$
The covariance matrix $\Omega_t$ is modelled with a discount factor
$\delta$ and $\Sigma$ is estimated following Triantafyllopoulos and
Pikoulas (2002) and Triantafyllopoulos (2006).

Table \ref{table8} shows the MSSE and the MAPE of the three
models. We see that all models produce reasonable results. For the
MSSE the best model is the TVVAR(2) (with the exception of the
lead variable where the MV-DLM produces MSSE closer to 1). For the
MAPE the best model is the DWR with the TVVAR(2) producing the
highest MAPE. Out of the three models, the MV-DLM is limited by
its mathematical form, which is constructed to give conjugate
analysis (see also the Introduction). The DWR suffers from similar
limitations as the MV-DLM, but it provides good results, for
linear trend time series without seasonality. The TVVAR model
provides a good modelling alternative and considering the numerous
applications of VAR time series models in econometrics, it is
believed that the TVVAR has a great potential.

In conclusion, the TVVAR model can produce forecasts with good
forecast accuracy, while the correlation of the series can be
estimated on-line with a fast linear algorithm. A criticism of the
model is that its efficiency depends on its order and if high order
TVVAR models are required (e.g. as in approximating moving average
processes with time-dependent parameters) its efficiency will be
similar of that of a vector MA, since the discount factor will have
to be close to 1. It will be interesting to know how the order of
the TVVAR model is related to the boundness of the eigenvalues of
the covariance estimator $\widetilde{S}_t$.

\section*{Concluding Comments}

This paper develops an algorithm for covariance estimation in
multivariate conditionally Gaussian dynamic linear models, assuming
that the observation covariance matrix is fixed, but unknown. This
is a general estimation procedure, which can be applied to any
Gaussian linear state space model. The algorithm is empirically
found to have good performance providing a covariance estimator
which converges to the true value of the observation covariance
matrix. The proposed methodology compares well with a non-sequential
state of the art MCMC estimation procedure and it is found that the
proposed estimates are close to the estimates of the MCMC. The new
algorithm is applied (but not limited to) model subclasses of VAR
and VAR with time-dependent parameters (TVVAR), which have great
application in financial time series. Considering the London metal
exchange data, it is found that the TVVAR model has outstanding
performance as opposed to the VAR model. It is believed that the
development of the TVVAR model is a worthwhile project and the
proposed fast, on-line algorithm for the estimation of the
observation covariance matrix, is a step forward opening several
paths for practical forecasting.

The focus in this paper is on facilitating and advancing
non-iterative covariance estimation procedures for vector time
series. Such procedures are particularly appealing, because of their
simplicity and ease in use. For such wide class of models such us
the conditionally Gaussian dynamic linear models, the proposed
on-line algorithm enables the computation of the mean vector of
standardized errors as well as it enables the computation of the
multi-step forecast covariance matrix. Both these computations are
valuable considerations in forecasting and they attract interest by
academics and practitioners alike.

\section*{Acknowledgements}

I should like to thank M. Aitkin and A. O'Hagan for helpful
discussions and suggestions on earlier drafts of the paper. Special
thanks are due to G. Montana who helped on the computational part of
the paper, in particular regarding the MCMC design and
implementation.

\renewcommand{\theequation}{A-\arabic{equation}} 
\setcounter{equation}{0}  

\section*{Appendix}  

\subsection*{Proof of Theorem \ref{th1}}

Let $\textrm{vech}(\cdot)$ denote the column stacking operator of
a lower portion of a symmetric square matrix and let $\otimes$
denote the Kronecker product of two matrices. First we prove that
for large $t$, it is approximately
\begin{equation}\label{eq4}
\E(\Sigma-\mathcal{A}_te_te_t'\mathcal{A}_t'|y^t)=\E(\Sigma-\mathcal{A}_te_te_t'
\mathcal{A}_t'|y^{t-1}),
\end{equation}
where $\mathcal{A}_t=n_t^{-1/2}S_{t-1}^{1/2}Q_t^{-1/2}$. Conditional
on $\Sigma$, we have from an application of the Kalman filter that
$\textrm{Cov}(e_{it}e_{jt},e_{kt}e_{\ell t}|\Sigma,y^{t-1})$ is
bounded, where $e_{t}=[e_{1t}~e_{2t}~\cdots ~e_{pt}]'$. Since
$\Sigma$ is bounded, $S_t$ is also bounded
($\lim_{t\rightarrow\infty}S_t=\Sigma$), and so all the covariances
of $e_{it}e_{jt}$ and $e_{kt}e_{\ell t}$ unconditional on $\Sigma$
are also bounded. This means that all the elements of
$\textrm{Var}\{\textrm{vech}(e_te_t')\}$ are bounded and so
$\textrm{Var}\{\textrm{vech}(e_te_t')\}$ is bounded. Now let
$$
\mathcal{X}_1=\E(\Sigma-\mathcal{A}_te_te_t'\mathcal{A}_t'|y_t,
y^{t-1})=S_t-\mathcal{A}_te_te_t'\mathcal{A}_t
$$
and
$$
\mathcal{X}_2=\E(\Sigma-\mathcal{A}_te_te_t'\mathcal{A}_t'|
y^{t-1})=S_{t-1}-\mathcal{A}_tQ_t\mathcal{A}_t'.
$$
Then, since $\lim_{t\rightarrow\infty}S_t=\Sigma$, there exists
appropriately a large integer $t(L)>0$ such that for every $t> t(L)$
it is $\E(\mathcal{X}_1-\mathcal{X}_2|y^{t-1})\approx 0$. Also
$$
\textrm{Var}\{\textrm{vech}(\mathcal{X}_1-\mathcal{X}_2)|y^{t-1}\}=\textrm{Var}\{(\mathcal{A}_t
\otimes\mathcal{A}_t)D_p\textrm{vech}(e_te_t')|y^{t-1}\}=\frac{1}{n_t^2}E_t\rightarrow
0,
$$
with
$$
E_t=\left[S_{t-1}^{1/2}Q_t^{-1/2}\otimes
S_{t-1}^{1/2}Q_t^{-1/2}\right]D_p
[\textrm{Var}\{\textrm{vech}(e_te_t')|y^{t-1}\} ] D_p'
\left[Q_{t}^{-1/2}S_{t-1}^{1/2}\otimes
Q_t^{-1/2}S_{t-1}^{1/2}\right],
$$
where $D_p$ is the duplication matrix and from the first part of the
proof we have that $E_t$ is bounded. It follows that for any $t>
t(L)$ it is $\mathcal{X}_1\approx \mathcal{X}_2$ with probability 1
and so we have proved equation (\ref{eq4}). Using
$\E(\Sigma|y^t)=S_t$, from equation (\ref{eq4}) we have
\begin{gather*}
\E (\Sigma|y^t)-\mathcal{A}_te_te_t'\mathcal{A}_t'=\E
(\Sigma|y^{t-1})-\mathcal{A}_t\E (e_te_t'|y^{t-1})\mathcal{A}_t' \\
\Rightarrow
S_t=S_{t-1}+\frac{1}{n_t}S_{t-1}^{1/2}Q_t^{-1/2}(e_te_t'-Q_t)
Q_t^{-1/2} S_{t-1}^{1/2} \\ \Rightarrow S_t =
S_{t-1}-\frac{1}{n_t}S_{t-1} + \frac{1}{n_t} S_{t-1}^{1/2}
Q_t^{-1/2}e_te_t'Q_t^{-1/2}S_{t-1}^{1/2}
\\ \Rightarrow n_tS_t=n_{t-1}S_{t-1}+ S_{t-1}^{1/2}
Q_t^{-1/2}e_te_t'Q_t^{-1/2}S_{t-1}^{1/2} = n_0S_0 +\sum_{i=1}^t
S_{i-1}^{1/2}Q_i^{-1/2}e_ie_i'Q_i^{-1/2}S_{i-1}^{1/2}
\end{gather*}
and by dividing by $n_t=n_0+t=n_{t-1}+1$ we obtain equation
(\ref{eq3}) as required.

\subsection*{The Gibbs Sampler for Multivariate Conditionally Gaussian DLMs}

The following procedure applies to any conditionally Gaussian
dynamic linear model in the form of equation (\ref{model1}). For the
simulation studies considered in this paper, given data
$y^N=(y_1,y_2,\ldots,y_N)$, we are interested in sampling a set of
state vectors, $\theta_1,\theta_2 \ldots, \theta_N$ and the
observation covariance matrix $\Sigma$ from the full, multivariate
posterior distribution of $\theta_1,\theta_2, \ldots, \theta_N,
\Sigma|y^N$.

Gibbs sampling involves iterative sampling from the full conditional
posterior of each $\theta_t, | \theta_{-t}, \Sigma, y^N$, for all
$t=1,2,\ldots,N$, and $\Sigma | \theta_1,\theta_2, \ldots, \theta_N,
y^N$; in our notation, $\theta_{-t}$ means that we are conditioning
upon all the components $\theta_1,\theta_2,\ldots,\theta_N$ but
$\theta_t$. Given the conditionally normal and linear structure of
the system, such full conditional distributions are standard, and
therefore easily sampled. However, such an implementation of the
Gibbs sampler, where each component is updated once at a time, could
be very inefficient when applied to the multivariate DLMs discussed
in this paper; in fact, the high-correlation of the dynamic system
will most likely bring convergence problems. In order to overcome
such difficulties, following the early suggestions of Carter and
Kohn (1994) and Fr\"{u}hwirth-Schnatter (1994), we have chosen to
implement a \emph{blocked} Gibbs sampler Gamerman (1997, p. 149);
within this context, this sampling scheme is better known as the
\emph{forward filtering, backward sampling} algorithm. Following is
a concise description of the algorithm used in our studies; for more
details, the reader should consult the references above, as well as
West and Harrison (1997, Chapter 15).

The first step of the Gibbs sampler involves sampling from the
updating distribution of $\theta_N | \Sigma, y^N$, which is given by
the multivariate normal $\mathcal{N}_d(m_N^M, P_N^M)$. This is done
in the forward filtering phase of the sampler, as follows. Starting
at time $t=0$ with some given initial values $m_0^M$, $P_0^M$ and
$\Sigma$ we compute the following quantities at each time $t$, for
$t=1,2, \ldots, N$:
\begin{itemize}
\item[(a)] the prior mean vector and covariance matrix of
$\theta_t|\Sigma,y^{t-1}$,
\begin{gather*}
a_t = G m_{t-1}^M  \quad \textrm{and} \quad R_t^M = G P_{t-1}^M G'
+ \Omega.
\end{gather*}
\item[(b)] the mean vector and covariance matrix of the one-step
ahead forecast of $y_t|y^{t-1}$,
\begin{gather*}
y_{t-1}^M(1) = F' a_t \quad \mbox{and} \quad Q_t^M = F' R_t^M F +
\Sigma .
\end{gather*}
\item[(c)] the posterior mean vector and covariance matrix of
$\theta_t|y^t$,
\begin{gather*}
m_t^M = a_t + A_t^M e_t^M \quad \mbox{and} \quad P_t^M = R_t^M -
A_t^M Q_t^M (A_t^M)',
\end{gather*}
where $A_t^M = R_t^M F (Q_t^M)^{-1}$ is the Kalman
  gain and $e_t^M = y_t - y_{t-1}^M(1)$ is the one-step ahead forecast error vector.
\end{itemize}

An updated vector $\theta_N$ is thus obtained, and the filtering
part of the algorithm is completed. The backwards sampling phase
involves sampling from the distribution of $\theta_t|\theta_{t+1},
\Sigma, y^t$ at all times $t=N-1, \ldots, 1, 0$. Each of such
vectors is drawn from a multivariate normal $N_d (h_t, H_t)$,
where
\begin{gather*}
h_t = m_t^M+P_t^MG'(R_{t+1}^M)^{-1}(\theta_{t+1}-a_{t+1})\quad
\textrm{and} \quad H_t = P_t^M\{I_d-G(R_{t+1}^M)^{-1}GP_t^M),
\end{gather*}
with $I_d$ being the $d\times d$ identity matrix. At each time
$t$, we also compute $\epsilon_t^* = y_t - F' \theta_t$. Once the
backwards sampling phase is completed, we set
$$
\widehat{\Sigma} = N^{-1} \sum_{t=1}^N \epsilon_t^*
(\epsilon_t^*)'.
$$

Finally, with $n_0^M$ being the prior degrees of freedom and $S_0^M$
being the prior estimate of $\Sigma$, we sample from the full
conditional density of $\Sigma|\Theta ,y^N$, which is an inverted
Wishart distribution
$\mathcal{IW}_p(n_0^M+N+2p,N\widehat{\Sigma}_N+n_0^MS_0^M) $, whose
simulation is also standard. This concludes an iteration of the
Gibbs sampler.

\end{document}